# 'Soft-Oxo-Metalates' (SOMs): A Very Short Introduction

Soumyajit Roy[1,2]

[1]School of Chemistry and Materials Engineering, Changshu Institute of Technology, Changshu, Jiangsu, P.R. China.

[2]Eco-friendly Applied Materials Laboratory, Materials Science Center, DCS, Indian Institute of Science Education and Research, Kolkata, India.

 roy.soumyajit@googlemail.com , s.roy@iiserkol.ac.in.

**Abstract.**

The field of polyoxometalates (POMs)[1], in recent times, has entered a new arena of 'soft'-supramolecular interactions, crossing the molecular regime of covalent bonds. The structures resulting from such 'soft'-supramolecular interactions are much larger (~10-500 nm in diameter) showing soft-matter properties. This comment proposes to name them 'Soft' Oxometalates. The comment after introducing and archiving seminal examples of Soft Oxometalates, gives few examples from the author's work where Soft Oxometalates are generated spontaneously and those designed by deliberate chemical reactions (also followed by very few selected examples from the literature). The comment ends with an open question: are all POMs soft?

# 1. Introduction

Ever since the publication of the seminal work by Müller[2] and later by Liu[3] an array of polyoxometalate (POM) structures at the border of translational periodicity has been discovered that calls for attention. Crossing the limits of translational periodicity these structures share some unique physical properties. For instance, they scatter light, possess a diffuse and mobile boundary, and have two phases a dispersed and a dispersive phase, show responsiveness to the change of dielectric constant of the dispersing media. Owing to such properties reminiscent of soft-matter, following de Gennes pioneering definition of soft-matter[4], this micro-review proposes to call these 'soft' structures of POMs as Soft Oxometalates (SOMs). [Note: These structures may also be called ‚Soft Polyoxometalates', which would be synonymus to the name of Soft Oxometalates (SOMs) which is used in this review. The author however believes the name „Soft Oxometalates" is more generic than „Soft Polyoxometalates". Reason: the basic oxometalate building blocks. The structures reported in this comment are all (mostly) super-structures of metal-oxides with a negative charge or oxometalates.[1a] The name 'polyoxometalates' however was historically coined to indicate single molecular structures with 'many' oxometalates.[1a] The present review sets out to give a name to a burgeoning field where the super-structures are primarily oxo-metalates in a supramolecular sense of the term. Hence although the name polyoxometalates (POMs) is more appropriate to refer to the structures at more single molecular regime, soft-oxometalates (SOMs) will describe oxometalate super-structures more at supra-molecular and soft polymer-colloidal length-scale regime.] This very small account (with an apology to many other active workers in this very diverse and active field) highlights few examples of relatively recent works and those of author's own which may be classified as Soft Oxometalates. The Soft Oxometalates may further be grouped in two categories: Spontaneously Formed Soft Oxometalates (SoFoSOMs) and Designed Soft Oxometalates (DeSOMs). Each category will be discussed with examples. (This micro-overview however does not give a recent state of the art description of the fascinating field of POMs. Interested readers may refer to reference [1].)

## 2. Spontaneously Formed Soft Oxometalates (SoFoSOMs):

Early works on SOMs can be traced back to the times of Berzelius[5] till the genius of Müller today to solve some of the long standing problems of the field. Why is the colour of Idaho Spring blue? Or, why is molybdenum blue, blue? A systematic seminal work by Müller's Bielefeld group revealed that it is the presence of a dynamic library of mixed valent Mo-based clusters whose Mo(V)→Mo(VI) IVCT band gives rise to such colour.[2a] They further first noticed the presence of colloidal length-scale particles in those solutions that in addition to colour, imparted a turbidity[2b] to such 'solutions'. It was then a combined effort of Liu group with Müller group that first delineated the structures of SOM-'blackberries' (or SoFoSOM-blackberries). Those are spherical entities comprising $\{Mo_{154}\}$ rings[3] and $\{Mo_{72}X_{30}\}$ type spheres, where X= Fe,[6a] $Mo_2$,[6b] W[6c]; held together by supramolecular interactions. The length-scales of such spheres are colloidal and hence scatter light. Several related SoFoSOMs have since then been isolated by Liu group.[7]

Recently a linear dependence between the radii of such SoFoSOM-blackberries with inverse of dielectric constant of the medium using $\{Mo_{72}Fe_{30}\}$ and $\{Mo_{132}\}$ as model systems was observed.[8] To explain such observation a simple model was proposed. This model proposes: 1. the driving force for the formation of SoFoSOM blackberries is pair-wise additive attractive interaction between the constituent POMs. 2. Their equilibrium size is determined by their renormalized charge density, which is in turn controlled by counter-ion condensation. Further from this model the interaction energy (cohesive/binding energy) that glues the units (each POM unit) of the SoFoSOM-blackberries together was calculated to be approximately 15 kJ/mol (at 300 K). This cohesive/binding energy is indeed 'soft' or supramolecular in the sense that it is comparable to the strength of a moderate X…H…X type hydrogen bond.[9] In other words these SoFoSOM blackberries may be justified to be called 'soft' not only because of their mobile, diffuse boundary but because of the 'soft' supramolecular interactions (whose energy is comparable in magnitude to that of moderate hydrogen bond) that glue the POM subunits in the final structure together.

Inspired by the work of Müller and Liu on molybdenum blue based Soft Oxometalates, another simple yet imminent question emerged. It is known that phosphomolybdate Keggins in aqueous suspensions form canary yellow precipitates. The question was: can the ammonium salt of phosphododecamolybdate Keggin, also show comparable SoFoSOM formation?[10] It is worth mentioning that phosphododecatungstate Keggin has been used in combination with AOT microemulsions or as templates to generate fibrous, star-like, and other interesting architectures on colloidal length scales even a Keggin based SOM microtube was reported.[11] Hence to address the above question of the nature of the colloidal objects in an aqueous dispersion of ammonium phosphododecamolybdate Keggin, we started our investigation with a very dilute sonicated dispersion of the salt. An investigation of sonicated aqueous colloidal dispersion of ammonium phosphododecamolybdate Keggin revealed first, the spontaneous formation of small spheres (5-50 nm radii). These spheres with time generate micrometer sized "pea-pod"-like particles. "Pea-pods" finally form rods (upon acidification). The whole phenomenon is shown schematically in Figure 1.

The above structures have been characterized by DLS, TEM, and STEM/EDX analyses. These analyses have revealed the decomposition of the starting phosphomolybdate Keggin along the lines of the formation of the peapods, responsible for the colloidal nature of the phosphomolybdate Keggin salt in water. (Hence it was proposed the phenomenon be called 'degenerative morphogenesis'.) It was further shown that low solubility and the presence of more than one component (like phosphate and molybdate in case of phosphomolybdate Keggin) in the starting precursor is necessary to observe such a phenomenon experimentally. It was thus demonstrated that less soluble polyoxometalates, such as ammonium phosphomolybdate (with two components, phosphate and molybdate), have the potential to form microscopic composites of SoFoSOMs (e.g., peapods), as observed experimentally. Thinking along these lines, it was predicted that Keggins could act as potential SoFoSOM precursors that could be candidates for forming comparable macroscopic composite SoFoSOM architectures.

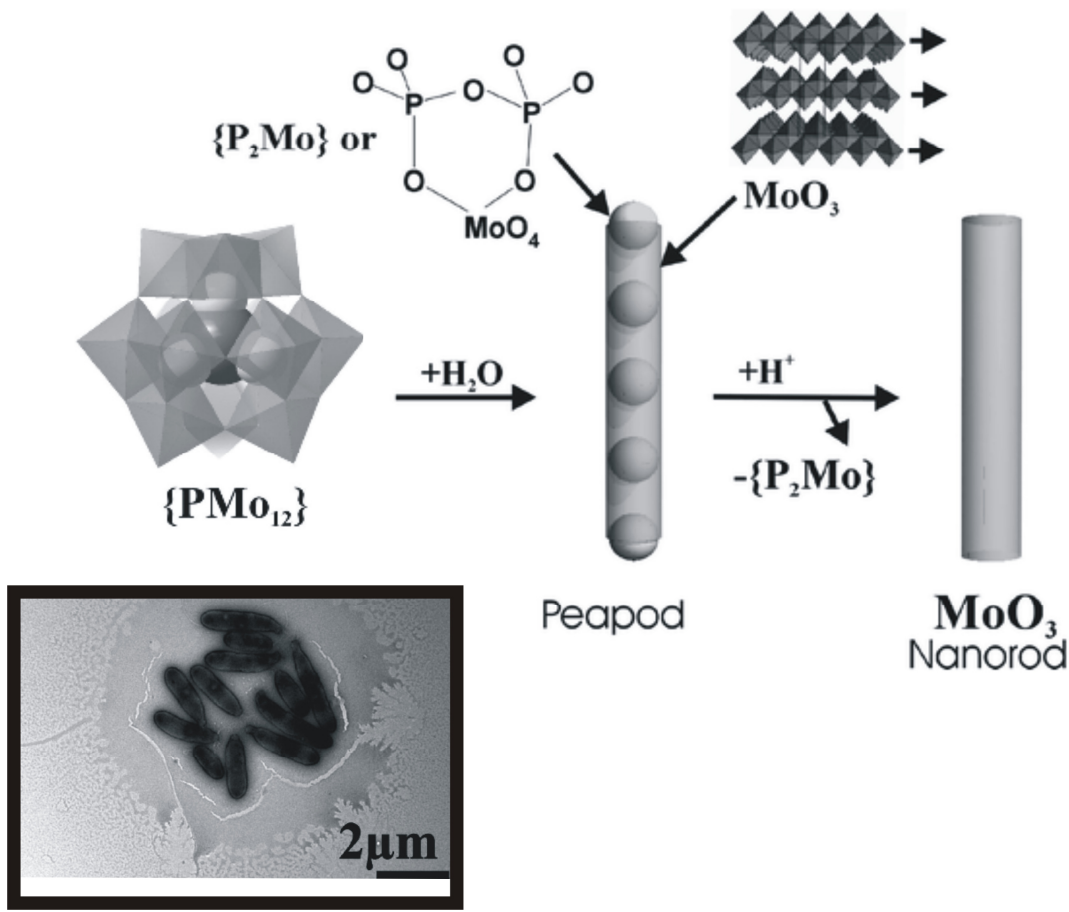

Figure 1. Spontaneously Formed Soft POM Peapod: Overview of the whole process of peapod formation from the starting ammonium salt of {PMo$_{12}$} Keggin. The spheres in the peapods are proposed to be composed of [P$_2$MoO$_{11}$]$^{6-}$, whereas the "skin" is proposed to consisting of a protonated MoO$_3$ sheet. A Bright Field TEM image of the SoFoSOM Peapods shown in the inset. For details please refer [10].

In a series of very elegantly designed experiments recently Lee group[12] has indeed shown that by using suitable couter-ion-Keggin based POM-precursor in water it is possible to show spontaneous generation of micrometer sized microtubular SoFoSOM formation in real time scale. They have attributed such structure formation to an osmotic pressure difference between the crystal phase and the solvent phase. It still remains interesting to be seen if Dawson-Wells, Anderson, Preyssler based POMs also form comparable SOM structures, and if any *a priori* pattern prediction for the final SoFoSOMs can be made based on the structure of the starting POM.

A note might be added to the concentration dependency for the generation of Soft Oxometalates at this point. As SoFoSOMs are dispersed structures of molecular POMs held together by soft supramolecular interactions, hence it is expected (and also observed) that they are found in solutions/dispersions of relatively low concentrations. Within this context, it is perhaps apt to mention that using Analytical Ultra Centrifugation (AUC) Sedimentation Velocity (SV) experiments, we were able to show the existence of single molecular {$Mo_{132}$} type Keplerates with acetate and sulfate as ligands, together with SoFoSOM blackberries, even at ultra-low concentrations.[13] Owing to the prevailing oxidative reaction conditions of the AUC-SV experiments and ultra-low concentration of Keplerates, extremely stable Keplerates, with acetates as ligands, were found to oxidatively open up to form {$Mo_{116}$} type open baskets; whereas the Keplerates with sulfate as ligands still remained closed, discrete. It is emphasized that extremely low concentration (~250 µmol) of the clusters coupled with an exposure to an oxidative reaction condition during the AUC-SV experiments (due to the presence of oxygen in the samples) leads to the low stability of the Keplerate {$Mo_{132}$} with acetate and is responsible for the slight deviation with respect to a recent report [Floquet, S; et. al. J. Am. Chem. Soc. 2009, 131, 888–903]. We further note that the unusual stability of the Keplerates at such low concentration and the oxidative opening of the cluster with acetate to form {$Mo_{116}$} type open basket is an intriguing property of the Keplerate with acetate as ligand at very low concentrations.[13]

### 3. <u>De</u>signed <u>S</u>oft <u>O</u>xometalates (DeSOMs):

Unlike the spontaneously generated SOMs that are held by soft, supramolecular interactions, there are numerous examples where such soft, supramolecular interactions have been employed to design Soft Oxometalates. This latter category can be proposed to be named as <u>De</u>signed <u>S</u>oft <u>O</u>xometalates (DeSOMs). Numerous design strategies have been employed; for instance, sol-gel method,[14] surfactant encapsulation,[15] Langmuir Blodgett method,[16] layer-by-layer technique,[17] solvent casting,[18] intercalation between layered hydroxides.[19] Exploiting electrostatic interactions between suitably charged colloidal templates/layered lattices/structured surfaces and POMs a class of DeSOMs have been synthesized. For instance, it might be mentioned on a passing that recently Wu group, has synthesized surfactant-encapsulated cluster (SEC)/polyoxometalate (SEP) type DeSOM complexes with terminal hydroxyl groups. The resulting DeSOMs – or SEC/SEPs with dangling hydroxyl groups have been further condensed in a silica matrix, using sol-gel process with tetraethyl orthosilicate (TEOS). The encapsulated POM structure inside the SEC/SEP DeSOMs hence remains unchanged. The formed hybrid, (strictly not a DeSOM but a ‚supra-DeSOM') has a well-defined hydrophobic nano-environment inside the DeSOM- or SEC/SEP, conducive to catalysis. [20] Beyond such DeSOMs formed by trapping of POM units by sol-gel method, they have also employed surfactant encapsulation as a route to generate POM containing gels or DeSOM-gels. [21] However originally, anisotropic DeSOM gels were first synthesized by Hasenknopf group using a Mn-$Mo_6$ type POM complex with bipridine and $Pd^{2+}$. [22] More recently, Carraro and Kortz group further showed, that POMs themselves could even actively initiate DeSOM formation by acting as gelators. For instance, they have shown that on addition of alcohols (EtOH, iPrOH, tBuOH, etc.) to an acidified aqueous solution of chiral $\{(R^*PO_3)Mo_5O_{15}\}$ type POMs gelation occurs. The gel network shows presence of 20 – 40 nm fibres with substructure consistent with H-bonded DeSOM network. [23] More recently, using Mn-$Mo_6$ type POMs with two dendritic poly(urethane amide) wings Nankai group showed interesting DeSOM formation that can act as pH sensitive gelators and form hybrid organogels of self-assembled hybrid nano-ribbons. [24] Apart from the above class of DeSOMs, it is to be noted that a body of venerable literature exploring catalytic acitivity of POMs has time and again employed diverse chemical means in addition to those as mentioned above for designing high surface area DeSOMs and the methods have been used for heterogenization of POM catalysts.[25] For instance, using, Rh(0), Ir(0), [26] Au(0)[27] clusters, Silica,[28] MOFs,[29] dendrimer polyelectrolytes,[30] as support DeSOMs have been reported and in many cases their catalytic activity documented. In principle, all such heterogenized POMs (even those which are employed with super-critical $CO_2$[31]) can come under the umbrella of DeSOMs. The incentive to synthesize such 'supported POMs' with a large surface area that could act as a 'bridge' between surface type catalysts and 'pseudoliquid phase' of bulk type catalysts remains high.[25d] Likewise, the question as of how to obtain such high-surface area POMs in a controlled way in an aqueous solution, given the fact that the surface area of molecular POMs are small and they tend to decompose[1] or superstructure to form SoFoSOMs (as mentioned in the previous section) in solution, remains relevant and important.

The examples below employ DeSOM design strategies which exploit electrostatic interaction between a colloidal entity (used as a structure directing agent/template/cast) and the molecular POM unit with charge. For instance, using this strategy, it was possible to 'glue' smaller POMs like the phosphomolybdate Keggins on the surface of gibbsite platelets to form large surface hexagonal DeSOM platelets.[32a] (Figure 2) (Note: some phosphomolybdate Keggin salts are unstable in dilute aqueous dispersion, forming SoFoSOMs as mentioned above in the previous section and the surface area of the molecular Keggins is order of magnitude smaller as compared to when glued on gibbsite as shown here.) However, it is still a challenge to predict and control cluster size and morphology of POMs in the mesoscopic regime (in the range of 100-900 nm). The reasons are difficulty to maneuver the interplay of chemistry of multiple metal centers, pH, and redox state of the complete system. Hence, in this regime, techniques like the one shown here and those reported in the literature can be crucial, since it bypasses the complex chemical crossroad and resorts to the use of preformed colloidal entities as templates/scaffolds for an engineered design of a mesoscopic POM architecture. The technique is evidently supramolecular (in the sense that it involves supramolecular electrostatic and induced dipole interactions way 'beyond the chemistry of molecules'), and it also provides a platform to 'glue' molecules to form mesoscopic supramolecular architecture or the DeSOMs. We further proposed to refer to this technique of using a colloidal particle as a template for forming large surface area DeSOMs as colloidal casting. The requirements for successful colloidal casting might be summarized as follows. (1) Complementarities of charge between the colloidal templates and the POM to be templated (as mentioned above, here it is between the positively charged gibbsite platelets and anionic Keggins); (2) a common solvent for both the components for effective templation (here water), and (3) formation of a stable colloidal dispersion. Likewise, it was possible further to test the applicability of the concept by changing the POM from phosphomolybdate Keggin to $\{Mo_{72}Fe_{30}\}$.[32b] As before, the complementarity of charge between $\{Mo_{72}Fe_{30}\}$ and gibbsite further prompted this choice. This charge complementarity can be understood as follows. The pH of discrete $\{Mo_{72}Fe_{30}\}$ clusters upon dissolution in water is around 4.5 whereas the isoelectric point or rather the point of zero charge of gibbsites is quite high, around pH 10.1. Hence, at a pH of around 4.5, the surface of the gibbsite nanocrystal is expected to be positively charged, (based on the following equilibrium), which in turn could act as a suitable platform for the attachment of anionic $\{Mo_{72}Fe_{30}\}$ clusters:

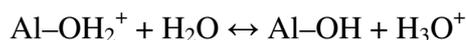

$$Al\text{–}OH_2^+ + H_2O \leftrightarrow Al\text{–}OH + H_3O^+$$

Consequently at a pH of 4.5, complementary charges on gibbsite nanocrystals and $\{Mo_{72}Fe_{30}\}$ clusters could act as glue for binding the latter on the surface of the former to form hexagonal platelets of $\{Mo_{72}Fe_{30}\}$ clusters, the basic strategy of this synthesis, same as before.

In another example, the topology of the template was changed, from hexagonal plates to spheres. In that example, a prefabricated cationic vesicle is used as a scaffold/structure directing agent and glues simple anionic oxomolybdates by electrostatic interaction and hydrogen bonds to form large DeSOM spheres.[33] (Figure 2) By this method, complexity in the resulting structure can be deliberately induced either via the scaffold or via the oxometalate. The high degree of control in the matter of the size and morphology of the resulting DeSOM superstructures renders this method attractive from a synthetic standpoint. For instance, the option of varying the overall DeSOM topology just by changing the shape of the vesicle exists and can still be explored. The synthesis was carried out by adding an appropriate amount of heptamolybdate to a previously prepared DOTAP vesicle (following standard practice). Interestingly, there is a window of heptamolybdate/DOTAP (M/D) concentration for which a stable dispersion is formed. However, beyond this window, the dispersion becomes unstable and later stable again. The phenomenon of the formation of a stable-unstable-stable dispersion was followed by electrophoretic mobility measurements, and it points to the operation of a charge inversion mechanism as a function of varying M/D concentration which may be explained as follows.

DOTAP vesicles are positively charged. Upon addition of anionic molybdates to this dispersion, the positive charge on the vesicles is reduced and finally instability is induced in a certain concentration window (1.5 > M/D > 0.6). The dispersion becomes nearly neutral and aggregates. If all the added POMs reside at the vesicles, this instability should manifest at M/D = 0.16, but the much higher observed values indicate the presence of free heptamolybdates in the dispersion. Consequently, an additional amount of heptamolybdate is necessary for reaching the unstable regime. However, upon further addition of molybdates (10 > M/D > 3), the dispersion undergoes charge inversion and is now stabilized by negative charges. From this ratio, the interface structure of the synthesized DeSOM giant spheres can immediately be deduced. More precisely, these experiments show that the surface charge density of both the DOTAP vesicle and the composite (for M/D ≈ 3 and higher) is same but only has opposite signs (i.e., +5 $\mu m\ cmV^{-1}s^{-1}$ in DOTAPs and -5 $\mu m cmV-1\ s-1$ in the composite). On the other hand, a DOTAP molecule carries a unit positive charge, whereas heptamolybdate has a charge of -6. From our above experimental results (i.e., charge inversion at M/D ≈ 3 and higher), it follows that in the DeSoPOM for every three DOTAP molecules, there is only one heptamolybdate. This picture is further consistent with the surface area of DOTAP [34] and heptamolybdate[1] reported in the literature. So, the DeSOM supersphere is most likely a vesicular DOTAP covered with a heptamolybdate monolayer where every three DOTAP molecules bind with one heptamolybdate of the monolayer. All the above DeSOMs were characterized by using a multitude of different techniques together, such as cryo-TEM (Transmission Electron Microscopy), TEM/EDX (TEM with Energy Dispersive X-Ray analyses), ATR-IR (Attenuated Total Reflection-Infra Red), Raman spectroscopy, Static and Dynamic Light Scattering, Small Angle X-Ray Scattering, Electrophoretic mobility measurements, potentiometric titrations, etc.

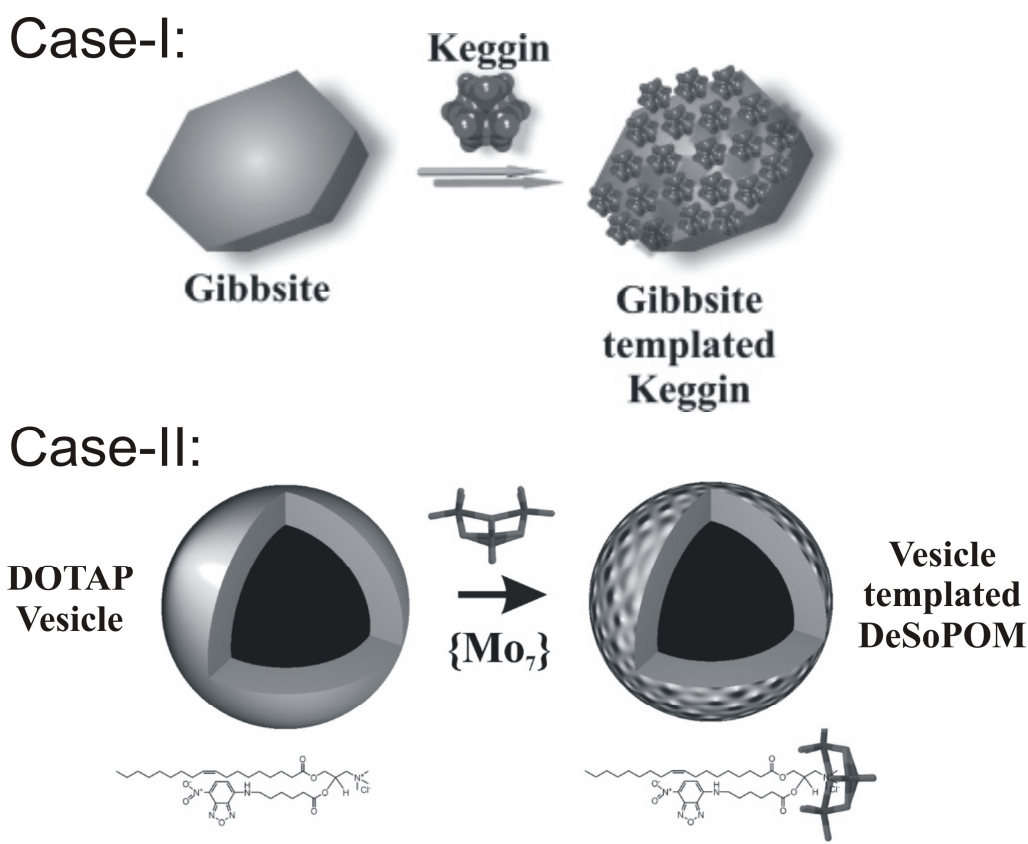

Figure 2. Two Cases of Designed Soft POMs: Case I: Schematic model of a gibbsite nanocrystal (left) used as a colloidal template for forming hexagonal mesoscopic DeSoPOM phosphomolybdate Keggin superstructure (right). The phosphomolybdate Keggin units are shown in a space filling model. (The picture is schematic only.) Case II: Schematic representation of the design strategy for making of POM large spheres. The spherical DOTAP template (left) and DOTAP-scaffold DeSoPOM supersphere (right) are shown schematically with surface corrugation. A plausible H-bonding between the DOTAP and the binding heptamolybdate (in wire-frame representation with Mo=O groups pointing toward the C=O) is implied. For details please refer [32] and [33].

Recently, Cronin and Liu group have reported with a very elegant set of experiments a comparable Mn-Anderson and surfactant based DeSOMs in mixed MeCN/$H_2O$ solvents.[35] In another very detailed work the kinetics and thermodynamics of DeSOM formation, using Alumino-,

Phospho-, and Silicotungstate exchanged with citrate stabilized gold nanoparticles by means of cryo-TEM and surface plasmon resonance study was demonstrated.[27] A tentative packing of the POM units on the DeSOM surface and the structural role of counter cations on the gold nanoparticles, in such DeSOMs was further proposed. In a slightly unrelated development *in silico* formation of DeSOM with Müller's {$Mo_{132}$} type Keplerate in lipid bilayers have also been shown.[36]

**4. Summary and Outlook:** To summarize along the above lines of ongoing discussions, it seems that there exists a structure-concentration continuum in the world of Oxometalates. The SOMs emerge at the lower concentration (more specifically, low volume fraction) whereas the crystalline POMs dominate the classical high concentration (high volume fraction regime). This view is lent from a work of Rosenbaum, Samora and Zukoski[37], where the authors have shown using second virial coefficients (osmotic compressibility data) precise regimes for crystallization of POMs using lithium, sodium salts of Silicotungstates as model systems. Hence it is perhaps worth saying crystalline POMs and SOMs though present in a 'virtual concentration phase space' manifest themselves only under suitable conditions of concentration/volume fraction, temperature, pressure and ionic strength. However the more interesting question still remains open: are all POMs soft? (Courtesy: Mobile anions of POM in the crystal lattice and related results and observations. [25d]) Perhaps future holds the answer.

**Acknowledgements:** The author thanks Changshu Institute of Technology, and Chair of the School of Chemistry and Materials Engineering, Changshu Institute of Technology, Jiangsu, China, and IISER-Kolkata, India for financial support.

**References:**

[1] (a) Pope, M. T. *Heteropoly and Isopoly Oxometalates*, Springer: Berlin, **1983**;

(b) Hill, C. L.; (Ed.) *Chem. Rev. Thematic Issue on Polyoxometalates*, **1998**, 98;

(c) Pope, M. T.; Müller, A. *Angew. Chem. Int. Ed. Engl.* **1991**, *30*, 34;

(d) Ginsberg, A. P.; (Ed.) *Inorganic Synthesis*, **1990**, *27,* 71;

(e) Cronin, L.; in McCleverty, J. A.; Meyer, T. J.; (Eds.) *Comprehensive Coordination Chemistry*, Elsevier, Amsterdam, **2004**, pp. 1;

(f) Yamase, T.; Pope, M. T.; (Eds.) *Polyoxometalate Chemistry for Nanocomposite Design*, Kluwer Academic Publishers, New York, **2002**;

(g) Pope, M. T.; Müller, A.; (Eds.) *Introduction to Polyoxometalate Chemistry : From Topology via Self-Assembly to Applications*, Kluwer Academic Publishers, New York, **2001**;

(h) Borra´s-Alamenar, J. J.; Coronado, E.; Müller, A.; Pope, M. T.; (Eds.) *Polyoxometalate Molecular Science, NATO Science series,* Kluwer Academic Publishers, Dordrecht, **2003**;

(i) Proust, A.; Thouvenot, R.; Gouzerh, P. *Chem. Commun.* **2008**, 1837.

[2] (a) Müller, A.; Serain, C. *Acc. Chem. Res.* **2000**, *33*, 2;

(b) Müller, A.; Diemann, E.; Kuhlmann, C.; Eimer, W.; Serain, C.; Tak, T.; Knöchel, A.; Pranzas, P. *Chem. Commun.* **2001**, 1928.

[3] (a) Liu, T.; Diemann, E.; Li, H.; Dress, A. W. M.; Müller, A. *Nature* **2003,** *426*, 59.

(b) Liu, T. *J. Am. Chem. Soc.* **2002**, *124*, 10942;

(c) Liu, T. *J. Am. Chem. Soc.* **2003**, *125*, 312;

See also: Zhu, Y.; Cammers-Goodwin, A.; Zhao, B.; Dozier, A.; Dickey, E. *Chem. Eur. J.* **2004,** *10*, 2421.

[4] de Gennes, P. G., *Soft Interfaces*, Cambridge University Press, **1997**.

[5] Berzelius, J. *Poggend. Ann. Phys. Chem.* **1826**, 6, 1206.

[6] (a) Zhang, J.; Li, D.; Liu, G.; Glover, K. J.; Liu, T. *J. Am. Chem. Soc.* **2009**, *131*, 15152;

(b) Kistler, M. L.; Bhatt, A.; Liu, G.; Casa, D.; Liu, T. *J. Am. Chem. Soc.* **2007**, *129*, 6453;

(c) Schäffer, C.; Merca, A.; Bögge, H.; Todea, A. M.; Kistler, M. L.; Liu, T.; Thouvenot, R.; Gouzerh, P.; Müller, A. *Angew. Chem. Int. Ed.,* **2009**, *48*, 149.

[7] Liu, G.; Liu, T.; Mal, S.; Kortz, U. *J. Am. Chem. Soc.* **2006**, *128*, 10103; Mishra, P. P. , Jing, J. , Francesconi, L. C., Liu, T. *Langmuir* **2008**, *24*, 9308; Zhang, J.; Keita, B.; Nadjo, L.; Martyr Mbomekalle, I.; Liu, T. *Langmuir* **2008**, *24*, 5277.

[8] Verhoeff, A.; Kistler, M.; Bhat, A.; Pigga, J.; Groenewold, J.; Klokkenburg, M.; Veen, S. J.; Roy, S.; Liu, T.; Kegel, W. K. *Phys. Rev. Lett.* **2007,** *99*, 066104.

[9] Jeffrey, G. A. *An Introduction to Hydrogen Bonding*, Oxford University Press, New York, **1997**.

[10] Roy, S.; Rijneveld-Ockers, M. T.; Groenewold, J.; Kuipers, B. W. M.; Meeldijk, H.; and Kegel, W. K. *Langmuir* **2007**, *23*, 5292.

[11] (a) Li, M.; Mann, S. *Langmuir* **2000**, *16*, 7088;

(b) Rautaray, D.; Sainkar, S. R.; Sastry, M. *Langmuir* **2003**, *19*, 10095;

(c) Mandal, S.; Rautaray, D.; Sastry, M. *J. Mater. Chem.* **2003**, *13*, 3002;

(d) Xin, Z.; Peng, J.; Wang, T.; Xue, B.; Li, L.; Wang, E. *Inorg. Chem.* **2006**, *45,* 8856.

[12] Ritchie, C.; Cooper, G. J. T.; Song, Y-F.; Streb, C.; Yin, H.; Parenty, A. D. C.; MacLaren, D. A.; and Cronin, L. *Nature Chemistry*, **2009**, *Vol. 1*, *April,* 47.

[13] Roy, S.; Planken, K. L.; Kim, R.; Mandele, D. v. d.; Kegel, W. K. *Inorg.Chem.* **2007**, *46*, 8469. See also: Roy, S.; Planken, K. L.; Kim, R.; Mandele, D. v. d.; Kegel, W. K. *Inorg.Chem.* **2010**, *49*, 5775.

[14] Polarz, S.; Smarsley, B.; Göltner, C.; Antonietti, M. *Adv. Mater.* **2000**, *12*, 1503.

[15] (a) Kurth, D.G.; Lehmann, P.; Volkmer, D.; Cölfen, H.; Koop, M.J.; Müller, A.; Du Chesne, A. *Chem. Eur. J.* **2000**, *6*, 385;

(b) Volkmer, D.; Du Chesne, A.; Kurth, D.G.; Schnablegger, H.; Lehmann, P.; Koop, M.J.; Müller, A. *J. Am. Chem. Soc.* **2000**, *122*, 1995.

(c) Kurth, D.G.; Lehmann, P.; Volkmer, D.; Müller, A.; Schwahn, D. *J. Chem. Soc. Dalton Trans.* **2000**, 3989.

[16] (a) Clemente-León, M.; Mingotaud, C.; Agricole, B.; Gómez-García, C. J.; Coronado, E.; Delhaès, P. *Angew. Chem. Int. Ed.* **1997**, *36*, 1114;

(b) Clemente-León, M.; Agricole, B.; Mingotaud, C.; Gómez-García, C. J.; Coronado, E.; Delhaès, P. *Langmuir* **1997**, *13*, 2340.

[17] Keller, S. W.; Kim, H.-N.; Mallouk, T. E. *J. Am. Chem. Soc.* **1994**, *116*, 8817.

[18] Song, I. K.; Kaba, M. S.; Coulston, G.; Kourtakis, K.; Barteau, M. A. *Chem. Mater.* **1996**, *8*, 2352.

[19] Kwon, T.; Pinnavaia, T. J. *J. Mol. Catal.* **1992**, *74*, 23.

[20] Qi, W.; Wang, Y.; Li, W.; Wu, L. *Chem. Eur. J.* **2010**, *16*, 1068.

[21] Wang, Y.; Li, W.; Wu, L. *Langmuir,* **2009**, *25*, 13194

[22] Favette, S.; Hasenknopf, B.; Vaissermann, J.; Gouzerh, P.; Roux, C. *Chem. Commun.* **2003**, 2664. For a recent review on metal and anion binding supramolecular gels see: Piepenbrock, M. M.; Lloyd, G. O.; Clarke, N.; Steed, J. W. *Chem. Rev.* **2010**, *110*, 1960.

[23] Carraro, M.; Sartorel, A.; Scorrano, G.; Maccato, C.; Dickman, M. H.; Kortz, U.; Bonchio, M. *Angew. Chem., Int. Ed.* **2008**, *47*, 7275

[24] Liu, B.; Yang, J.; Yang, M.; Wang, Y.; Xia, N.; Zhang, Z.; Zheng, P.; Wang, W.; Lieberwirth, I.; Kübel, C. *Soft Matter*, **2011**, *7*, 2317

[25] For reviews on POMs in catalysis, see:

(a) Kozhevnikov, I. V.; (Ed.) *Catalysts for Fine Chemicals Synthesis: Catalysis by Polyoxometalates,* Vol. 2, J. Wiley & Sons, Chichester, **2002**;

(b) Hill, C. L.; Prosser-McCartha, C. M. *Coord. Chem. Rev.* **1995**, *143*, 407;

(c) Mizuno, N.; Misono, M. *Chem. Rev.* **1998**, *98*, 199;

(d) Kozhevnikov, I. V. *Chem. Rev.* **1998**, *98*, 171;

(e) Rocchiccioli-Deltcheff, C.; Aouissi, A.; Bettahar, M.; Launay, S.; Fournier, M. J.; *J. Catal.* **1996**, *164*, 16;

(f) Neumann, R.; Khenkin, A. M. *Chem. Commun.* **2006**, 2529.

[26] Finke, R. G.; Özkar, S. *Coord. Chem. Rev.* **2004**, *248*, 135.

[27] Wang, Y.; Neyman, A.; Arkhangelsky, E.; Gitis, V.; Meshi, L.; Weinstock, I. A. *J. Am. Chem. Soc.* **2009**, *131*, 17412.

[28] (a) Okun, N. M.; Anderson, T. M.; Hill, C. L. *J. Am. Chem. Soc.* **2003**, *125*, 3194;

(b) Okun, N. M.; Ritorto, M. D.; Anderson, T. M.; Hill, C. L. *Chem. Mater.* **2004**, *16*, 2551.

[29] Sun, C.-Y.; Liu, S.-X.; Liang, D.-D.; Shao, K.-Z.; Ren, Y.-H,; Su, Z.-M. *J. Am. Chem. Soc.* **2009**, *131*, 1883. See also for a review: Corma, A.; García, H.; Llabrés i Xamena, F. X. *Chem. Rev.* **2010**, *doi: 10.1021/cr9003924*.

[30] (a) Plault, L.; Hauseler, A.; Nlate, S.; Astruc, D.; Ruiz, J.; Gatard, S.; Neumann, R. *Angew. Chem., Int. Ed.* **2004**, *43*, 2924;

(b) Nlate, S.; Astruc, D.; Neumann, R. *Adv. Synth. Catal.* **2004**, *346*, 1445.

[31] Maayan, G.; Ganchegui, B.; Leitner, W.; Neumann, R. *Chem. Commun.* **2006**, 2230.

[32] (a) Roy, S.; Mourad, M. C. D.; Rijneveld-Ockers, M. T. *Langmuir* **2007**, *23*, 399;

(b) Roy, S.; Meeldijk, H. J. D.; Petukhov, A. V.; Versluijs, M.; Soulimani, F. *Dalton Trans.* **2008**, 2861.

[33] Roy, S.; Bossers, L. C. A. M.; Meeldijk, H. J. D.; Kuipers, B. W. M.; Kegel, W. K. *Langmuir* **2008**, *24*, 666.

[34] Generosi, J.; Castellano, C.; Pozzi, D.; Congiu Castellano, A.; Felici, R.; Natali, F.; Fragneto, G. *J. Appl. Phys.* **2004**, *96 (11)*, 6839.

[35] Zhang, J.; Song, Y.-F.; Cronin, L.; Liu, T. *J. Am. Chem. Soc.* **2008**, *130*, 14408.

[36] Carr, R.; Weinstock, I. A.; Sivaprasadarao, A.; Müller, A.; Aksimentiev, A. *Nano Lett.* **2008**, *8(11)*, 3916.

[37] Rosenbaum, D.; Zamora, P. C.; Zukoski, C. F. *Phys. Rev. Lett.* **1996**, *76*, 150.